\documentclass[preprint,12pt]{elsarticle}

\usepackage{amsthm}
\usepackage{float}
\usepackage{bm}
\usepackage{subfig}
\usepackage{graphicx}
\usepackage{epsfig}
\usepackage{mathtools}
\usepackage{hyperref}
\usepackage{booktabs}
\usepackage{pgfplotstable}

\journal{Nuclear Physics B}

\begin{document}

\begin{frontmatter}



\title{Energy Consumption Analysis Of Machining Centers Using Bayesian Analysis And Genetic Optimization}


\author[inst1]{Johnatan Cardona Jiménez}
\affiliation[inst1]{organization={Faculty of Engineering, Pascual Bravo University},
            city={Medellin},
            country={Colombia}}

\author[inst1]{María I. Ardila}
\author[inst1]{J. S. Rudas}
\author[inst1]{Cesar A. Isaza M.}
\author[inst1]{Edwin J. Núñez}
\author[inst2]{Miguel A. Rodriguez}
\affiliation[inst2]{organization={Faculty of Engineering, Instituto Tecnologico Metropolitano},
            city={Medellin},
            country={Colombia}}

\begin{abstract}
Responding to the current urgent need for low carbon emissions and high efficiency in manufacturing processes, the relationships between three different machining factors (depth of cut, feed rate, and spindle rate) on power consumption and surface finish (roughness) were analysed by applying a Bayesian seemingly unrelated regressions (SUR) model. For the analysis, an optimization criterion was established and minimized by using an optimization algorithm that combines evolutionary algorithm methods with a derivative-based (quasi-Newton) method to find the optimal conditions for energy consumption that obtains a good surface finish quality. A Bayesian ANOVA was also performed to identify the most important factors in terms of variance explanation of the observed outcomes. The data were obtained from a factorial experimental design performed in two computerized numerical control (CNC) vertical machining centers (Haas UMC-750 and Leadwell V-40iT). Some results from this study show that the feed rate is the most influential factor in power consumption, and the depth of cut is the factor with the stronger influence on roughness values. An optimal operational point is found for the three factors with a predictive error of less than 0.01\% and 0.03\% for the Leadwell V-40iT machine and the Haas UMC-750 machine, respectively.
\end{abstract}




\end{frontmatter}


\section{Introduction}
\label{sec:1}
Energy-saving and emissions reduction are important challenges faced by different economic sectors due to their negative potential impacts on the environment and the acceleration of climate change. Improving energy efficiency, which reduces energy consumption, is an important task for almost all industrial processes, especially for the manufacturing industry. This poses a challenging opportunity to mitigate the current levels of emissions and reduce energy use through an appropriate assessment and characterization of manufacturing industry operations for improving their performance \citep{Cai2021}. According to the International Energy Agency (IEA), the manufacturing
industry is responsible for approximately 33\% of global energy consumption and as much as 36\% of global carbon emissions \citep{Cai2021,Sihag2020}. For machining processes (i.e., drilling, turning, or milling), a maximum efficiency of 30\% has been reported \citep{Sihag2020}. In general, the industrial sector is expected to gradually grow over the next few decades, leading to increased demand for machining processes and, consequently, increased energy consumption and carbon emissions \citep{Moradnazhad2017}. Energy savings for machining processes can be classified into two branches. The first is the improvement of cutting technologies, such as high-speed, cutting tools and their materials. The second is to examine the relationship between the process parameters and the power required for process functionality to optimize energy consumption by applying models and optimization methods for energy savings.  For instance, \cite{Li2017} proposed two alternatives to analyse the optimization problem: design strategies and operational strategies. Design strategies include all the aspects related to machine improvement,    such as power aspects and maintenance prediction, which includes the control systems \citep{Kordonowy2002,Pavanaskar2014}. For most common industries, design strategies are not the best solution because they require downtime or machine center inactivity for strategy implementation, which can increase economic losses. On the other hand,      operation strategies correspond to all the operational variables and tools, including cutting parameters, tool paths, and process planning \citep{Mukherjee2006, Zhou2016,Yan2013}. The most common factors considered in operational strategies are the feed rate \cite{Peng2013, Behrendt2012, Draganescu2003}, depth of cut \cite{Peng2013, Pavanaskar2014, Cai2021, Kordonowy2002, Abishekraj2020}, cutting speed \cite{KumarParida2019, Draganescu2003, Upadhyay2013} cutting edge radius  \cite{Edem2018, Fulemova2014} and vibration signals \cite{Chen2017, Upadhyay2013}. Operational strategies have been focused on output variable optimization through experimental, theoretical, or mixed approaches. Some of the most studied output variables reported in the literature are the specific energy coefficient or power machining consumption, the surface finish or surface roughness, and the material removal rate \citep{Peng2013, Pavanaskar2014, Cai2021, Kordonowy2002, Abishekraj2020, KumarParida2019, Draganescu2003,jmmp4030066,jmmp4030064}.
 
 Currently, engineers are aware that setting the correct operational variables is the manufacturer’s main purpose. Thus, truly understanding the interdependencies among the main operational factors and establishing the correct levels for the inputs in the machining process is a technological challenge \citep{KumarParida2019, Misaka2020, Manivel2016}. In this sense, several authors have developed and proposed different approaches for optimizing machining parameters. For example, \cite{KumarParida2019} developed a flank wear and surface roughness prediction model for a hot turning machining process based on response surface methodology (RSM). Their model had      an 86.17\% accuracy for      surface roughneslinearities. However, its prediction accuracy is not the best for all phenomena. Considering a different approach, \cite{Chen2017} used an artificial neural network (ANN) for surface roughness prediction considering the cutting speed, feed rate, and the depth of cut as input factors. The authors validated their hypothesis about the strong relationship between the feed rate and the surface finish, and they reported a maximum relative prediction error of 12.2\% for the ANN. \cite{BENARDOS2002343} proposed an ANN for the prediction of surface roughness. Their experiments were conducted on a CNC milling machine according to the principles of Taguchi’s design of experiments, where the depth of cut, the feed rate per tooth, the cutting speed, the engagement and the wear of the cutting tool, the use of cutting fluid, and the three components of the cutting force were the controlled parameters. The author found the most influential parameters and attained a prediction level for the surface roughness with a mean percentage squared error equal to 1.86\%. In a similar approach, \cite{GHANI200484} used a Taguchi optimization method for evaluating the cutting speed, feed rate, and cut depth of cut in AISI H13 steel machined with a TiN-coated P10 carbide tool. The authors used an orthogonal array, signal-to-noise (S/N) ratio, and Pareto analysis of variance (ANOVA) to analyse the effect of these milling parameters. They concluded that the Taguchi metho was viable for solving machining problems with a minimum number of procedures. \cite{TANDON2002595} used      an ANN predictive model for critical process parameters to predict the cutting forces. For its implementation, the authors used a particle swarm optimization (PSO) algorithm to optimize cutting conditions such as feed and speed. They reported machining time reductions of up to 35\%. \citep{GHOSH2007466} studied      the cutting forces, spindle motor current, and sound pressure level in a CNC milling process  to examine tool wear. The authors used a neural network-based sensor fusion model and proposed newer methods such as feature space filtering and prediction space filtering, among others, to improve prediction accuracy in a real-time error-prone environment.  \citep{Radhakrishnan2005} implemented another predicted cutting force model using regression and an ANN. The authors used the methodology to filter out abnormal data points, and the filtered data were used in a neural network for better prediction.

\begin{figure}[H]
\centering
\includegraphics[scale=0.5]{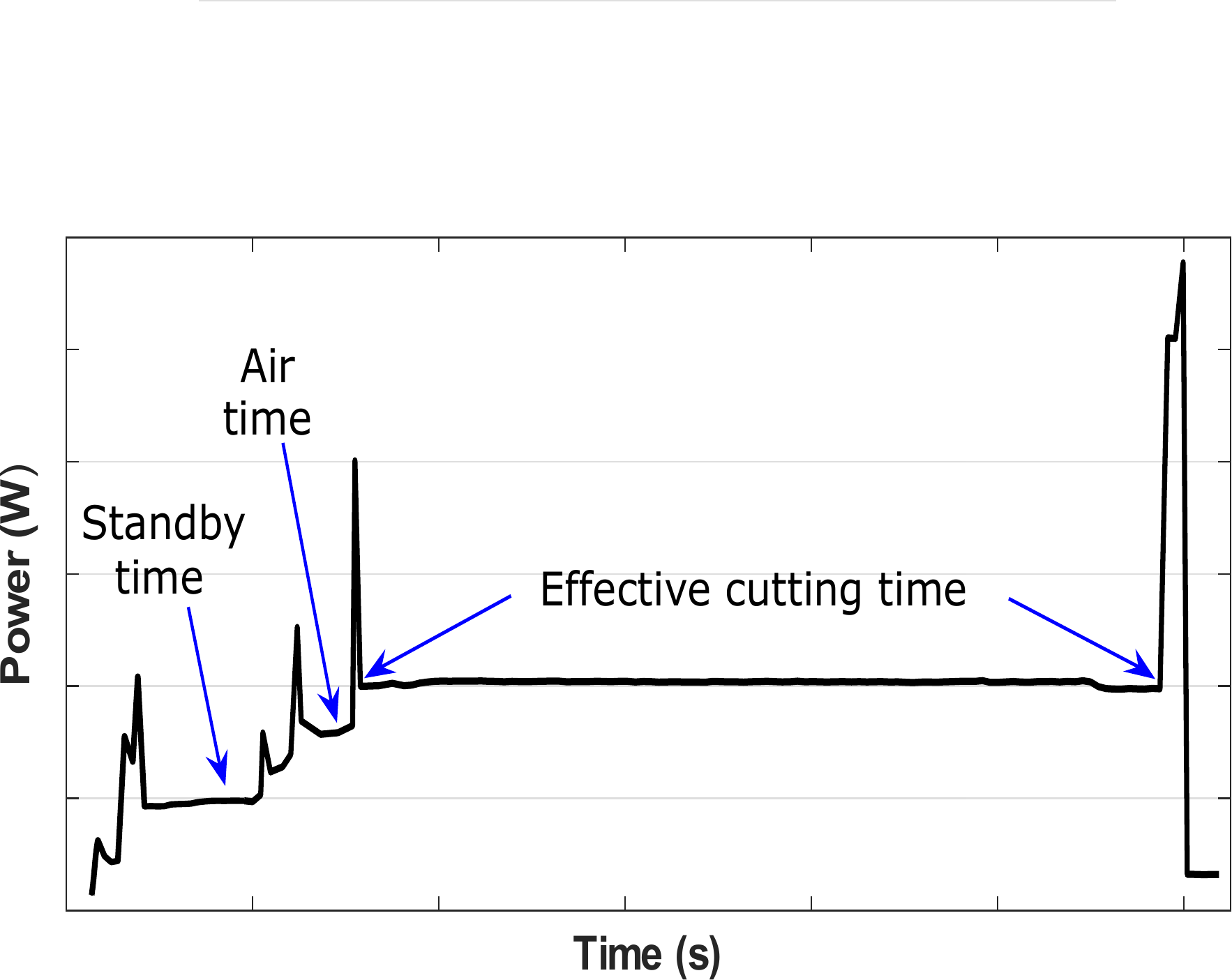} 
\caption{Power profile of a machining process}
\label{fig1_1}
\end{figure}

During a machining process, the cutting time is composed of different stages: the standby time (the whole system going towards the machining planning setup), the air time (the tool does not have any  applied load), and effective cutting time (material removal through tool movement). A schematic diagram of the power profile showing the cutting stages is illustrated in Figure \ref{fig1_1}. All these cutting stages      have a strong influence on the energy consumption for this type of process. Similarly, several studies have shown that the toolpath, cutting speed and depth, and spindle speed can strongly influence the effective cutting time. The effect of these parameters on the total power consumption of machining have also been noted \citep{Li2017, Manivel2016, Misaka2020}. Regardless of the known relationship between the process parameters, the power consumption, and the machining quality, it is necessary to extensively study, analyse, and predict the quality of the surface finish of manufactured parts considering optimal energy performance through the proper choice of operational factors.

This work presents a novel statistical analysis for an experimental study to analyse the energy consumption (EC) and surface roughness (SR) and their relationship with three controllable machining parameters. Specifically, in this experiment, two ma chine      centers (Haas UMC-750 and Leadwell V-40iT® CNC) are tested under different experimental conditions to characterize their EC and SR produced over experimental units both machines have processed. The three machine parameters related to the experimental conditions are depth of cut, feed rate, and spindle rate. Considering the plausible observed correlation between the experimental outcomes (EC and SR), a bivariate response regression linear model is used. A Bayesian approach is applied for the inferential analysis of the latent quantities related to the statistical modelling. An optimization criterion based on the joint minimization of both outcomes is defined to find the optimal operational conditions for both machining centers. In addition, a genetic algorithm is used to identify the experimental conditions that jointly minimize the EC and SR for both machining      centers. To characterize the relevance of each controllable machining parameter on both responses (EC and SR), a Bayesian version of the classic analysis of variance (ANOVA) is used. The next section describes the experimental setup employed in this study. It also presents the statistical model and optimization procedures used to analyse      the data obtained from the experimental runs. Section 3 presents the main results and analysis from this study. Finally, Section 4 provides some concluding remarks and potential open problems for future development.

\section{Materials and Methods}
\label{sec:2}
\subsection{Machining centers and experimental design}

The experimental runs are carried out on two CNC vertical machining centers: a Haas UMC-750 with a 12000 rpm maximum speed and 22.4 kW of spindle power and a Leadwell V-40iT with a 10000 rpm maximum speed and 18.5 kW of spindle power. For a matter of simplicity, from this point on, the Leadwell V-40 iT and the Haas UMC-750 machining      centers are referred to as machine A and machine B, respectively.      Carbon SAE 1045 steel with a hardness of 190 HB (experimental units) is utilized for the machining operation. The workpiece chemical composition is presented in Table \ref{tabla_1}. A Fluke 1732 3-phase electrical energy logger measures power consumption during the CNC machining processes. As mentioned above, the cutting time has different stages (see Figure \ref{fig1_1}). In this work, the energy consumption is obtained by integrating the electrical power and considering the time of the effective cutting interval for each experimental run. Additionally, the surface finish quality is evaluated using a Tmteck TMR200 portable surface roughness checker, measuring Ra ($\mu$m).  The cutting tools are master cut from 1/2'' coated tungsten carbide.

\begin{center}
\begin{table}[H]
\caption{Chemical composition of carbon steel SAE 1045}
\begin{center}
\begin{tabular}{cccccc}
\hline
\textbf{Element} & C         & Mo      & P   & S    & Si      \\ \hline
\textbf{Wt\%}    & 0.43-0.50 & 0.6-0.9 & 0.4 & 0.05 & 0.2-0.4 \\ \hline
\end{tabular}
\label{tabla_1}
\end{center}
\end{table}
\end{center}


A factorial design has been selected, and the controllable factors with their corresponding levels were chosen according to previous works  \cite{Li2017, Liu2019}. One of the main purposes of this design is to obtain the optimal machining setting conditions by considering the power consumption and roughness as response variables. The experimental setup is summarized in Table \ref{tabla_2}.

\begin{center}
\begin{table}[h]
\caption{Experimental setup.}
\begin{center}
\begin{tabular}{cccccccc}
\hline
 &  &  & \multicolumn{5}{c}{\textbf{Levels}}                             \\ \cline{4-8}  \textbf{Factors}
                                 &  \textbf{Units}                              &                            \textbf{Variable name}                                                    & \textbf{1} & \textbf{2} & \textbf{3} & \textbf{4} & \textbf{5} \\ \hline
Depth of cut                     & mm                             & $X_{1}$                                                                           & 1          & 1.5        & 2          & 2.5        &  3        \\ \hline
Feed rate                        & mm/min                         & $X_{2}$                                                                           & 134        & 167.5      & 201        & 234.5      & 268        \\ \hline
Spindle rate                     & RPM                            & $X_{3}$                                                                           & 950        & 1187.5     & 1425       & 1662.5     & 1900       \\ \hline
\end{tabular}
\label{tabla_2}
\end{center}
\end{table}
\end{center}

\begin{figure}[H]
\centering
\includegraphics[scale=1.5]{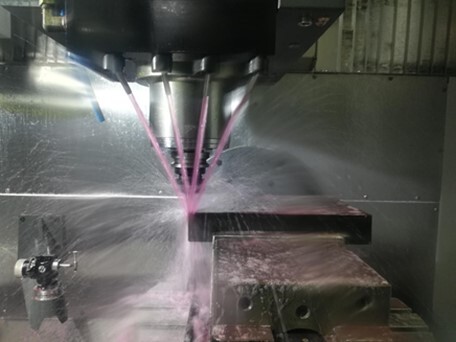}
\caption{Experimental setup for the machining process}
\label{fig_2}
\end{figure}

Given the number of combinations from the levels of the three factors presented in Table \ref{tabla_2},  125 experiments are defined per machining center. Thus, the total number of experimental runs is 250.  Some pilot experiments are carried out to set up the planning process, including the preprocessing and postprocessing algorithm for the Sprutcam software used during the machining process.  Figure \ref{fig_2} shows the setup for side milling experimentation.

\subsection{Statistical Analysis}

The two outcomes obtained from the experiment are related to the same experimental units; therefore, some significant level of correlation between them could be present.      Thus, to take into account this relationship in the modelling, the seemingly unrelated regressions (SUR) model has been employed \citep{zellner1962efficient, greene2000econometric} to relate the outcomes to the covariates or fixed factors. The specification of the model is as follows:

\begin{equation}\label{equ1}
\begin{split}
log(\textbf{y}_i) &= \beta_{i0} +  I(m)\beta_{i7} + \sum_{j=1}^{3}( X_{j}\beta_{ij} + X_{j}^2\beta_{i,3+j} + X_{j}I(m)\beta_{i,j+7}) + X_1X_2\beta_{11} +\\
&\quad X_1X_3\beta_{12} + X_2X_3\beta_{13} + \epsilon_i,
\end{split}
\end{equation}for $i=1,2$. Thus, $i=1$ and $i=2$, when the observed outcome is the roughness ($\textbf{y}_1$) and the power ($\textbf{y}_2$), respectively. $X_1$ (depth of cut), $X_2$ (feed rate), $X_3$ (spindle rate), represent the controlled factors, and $I(m)$ is a dummy variable that defines a set of equations for each machining center that has been evaluated. In this way, we have 

\begin{equation*}
  I(m) =
    \begin{cases}
      1 & \text{if it is the machine}\ \ m=B\\
      0 & \text{if it is the machine}\ \ m=A,
    \end{cases}       
\end{equation*}which consequently leads to

\begin{equation*}
  log(\textbf{y}_i) =
    \begin{cases}
      \beta_{i0} + \beta_{i7} + \sum_{j=1}^{3}( X_{j}(\beta_{ij} + \beta_{i,j+7}) + X_{j}^2\beta_{i,3+j}) + X_1X_2\beta_{11} +\\
X_1X_3\beta_{12} + X_2X_3\beta_{13} + \epsilon_i, & \text{for}\ \ m=B\\
      \beta_{i0} + \sum_{j=1}^{3}( X_{j}\beta_{ij} + X_{j}^2\beta_{i,3+j}) + X_1X_2\beta_{11} + X_1X_3\beta_{12} +\\ X_2X_3\beta_{13} + \epsilon_i & \text{for}\ \ m=A.
    \end{cases}       
\end{equation*}

The term $\epsilon_i$ is the random error related to all known and unknown factors that have not been controlled in the experiment, and therefore not considered in the deterministic component in the model (\ref{equ1}). It is assumed that $\epsilon_i \sim N(0, \sigma_i I_n)$, which means that the random error is normal distributed with mean zero and constant variance  $\sigma_i$, where $I_n$ is the $n\times n$ identity matrix. As we aim to model both equations jointly, the vector $\bm{\epsilon} = (\epsilon_1, \epsilon_2)^{'}$ is defined and it is assumed that $\bm{\epsilon}\sim N(0, \Sigma)$, where $\Sigma$ is the covariance matrix that accounts for the individual variances related to each outcome and the covariance structure between them. Given that a Bayesian approach is being followed and there is no prior information available for this experiment, non-informative or vague prior normal distributions are set for the $\beta$ parameters and an inverse Wishart distribution is set for the $\Sigma$ parameter. There are plenty of possible justifications for the adoption of a Bayesian approach; as such, we highlight the modelling flexibility it offers by considering the parameters as random variables, the option to include prior information (prior to the experiment) whenever available, and the advantage that its inferential properties hold for any sample size, even small samples. For a better understanding of the Bayesian paradigm, the interested reader is referred to \cite{gelman2013bayesian} and \cite{o1994kendall}. To assess the significance of the parameters in the model (\ref{equ1}), high-density intervals (HDI) are used, which are the Bayesian counterpart of the confidence intervals (CI).

\subsection{Optimal desing}
Based on model (\ref{equ1}), we aim to find the experimental conditions $E = \{X_1, X_2, X_3\}$, which would produce an optimal outcome. What an optimal response could mean depends on the interest of the experimenters. As mentioned above, in this work, the interest is to find the experimental conditions that minimize both the energy consumed (power) by the two machining centers and the roughness value generated over the experimental units. To do so, we define the minimum vector as $Y^{min}=(y_1^{min}, y_2^{min})$, where $y_1^{min}$ and $y_2^{min}$ are the minimum thresholds (for energy and roughness, respectively) the experimenters are seeking to attain. Thus, within this framework, the optimal design $E^*=\{X_1, X_2, X_3\}$ is defined as follows:

\begin{equation}\label{funcM}
d(E^*)=\min_{\forall E \in \mathcal{E}} d(\hat{Y},Y^{min}),
\end{equation}where $\mathcal{E}$ is the set of all possible designs within the region where the experiment is performed, $\hat{Y}$ is the predicted response obtained from model (\ref{equ1}), and $d(\cdot,\cdot)$ is a metric or distance function. To minimize the function $d(\cdot,\cdot)$, we apply genetic optimization using derivatives \citep{GeneticOptim1}, specifically the version implemented in the \textbf{R} package rgenoud \citep{GeneticOptim2, Rcite}. gnoud is the package's function that combines evolutionary algorithm methods with a derivative-based (quasi-Newton) method to solve difficult optimization problems. It is also possible to use this function for optimization problems in which derivatives do not exist. It can also be applied to optimize nonlinear or discontinuous functions in the variables of the function to be optimized. One of the main advantages of using this method is that the genetic optimizer rarely gets stuck at regions of local minimum points.

\subsection{Bayesian ANOVA}
ANOVA is a popular method widely used in data analysis related to experimental designs. In this work, we employ the Bayesian ANOVA analysis proposed by \cite{gelman2005analysis}. It can be viewed as a graphical tool, that allows identification of the more relevant factors in explaining or predicting the variability of an observed response. This Bayesian version of the ANOVA is also based on a fixed/random-effects model, which in this case is defined by: 

\begin{equation}\label{anova2}
y_{ij} = \mu + \alpha_j^{D} +  \alpha_j^{A} + \alpha_j^{V} + \epsilon_{ij}, 
\end{equation}where $\mu$ is the global mean for the whole experiment, $\alpha_j^{f}$ represents the $j$-th treatment effect related to each factor controlled in the experiment ($f=\{D:\text{depth of cut}, A:\text{feed rate}, V:\text{spindle rate}\}$), for $j=1,\ldots,k_f$, where $k_f$ is the number of levels of factor $f$. Thus, the finite population variance is given by

\begin{equation}
S_{\bm{\alpha}^{f}}= \sqrt{ \frac{1}{k_f-1}\bm{\alpha}^{'f}\left[\textbf{I}_{k_f} - \frac{1}{k_f}\textbf{J}\right]\bm{\alpha}^{f}},
\end{equation}where $\bm{\alpha}^{f}=(\alpha_1^{f}, \ldots, \alpha_{k_f}^{f})^{'}$, $\textbf{I}_{k_f}$ is the $k_f\times k_f$ identity matrix, and $\bm{J}$ is the $k_f\times k_f$ matrix of ones. As stated by \citep{kaufman2010bayesian}, the traditional mean square can be thought of simply as a point estimate of this quantity. In contrast, the Bayesian approach provides a full posterior distribution for $S_{\bm{\alpha}^{f}}$. Thus, the controlled factors with the higher $S_{\bm{\alpha}^{f}}$ values will be the variables that are more relevant for the explanation of the variability observed in both response variables. The factors with lower or close to zero $S_{\bm{\alpha}^{f}}$ values can be interpreted as irrelevant variables for the analysis. This analysis is used to assess the relative importance of the (random or fixed) effects in a Bayesian hierarchical linear model rather than testing their significance.

\section{Results and discussion}
\begin{figure}[H]
\centering
\subfloat []{\includegraphics[scale=1.45]{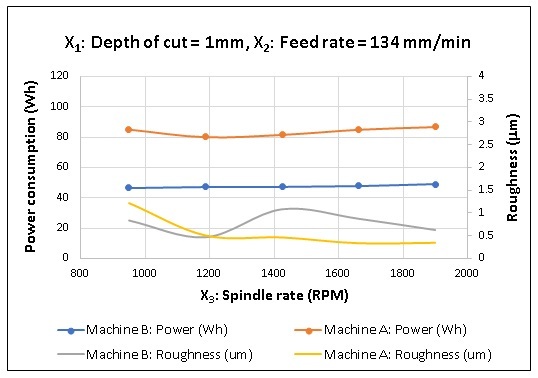}}
\subfloat[]{\includegraphics[scale=1.45]{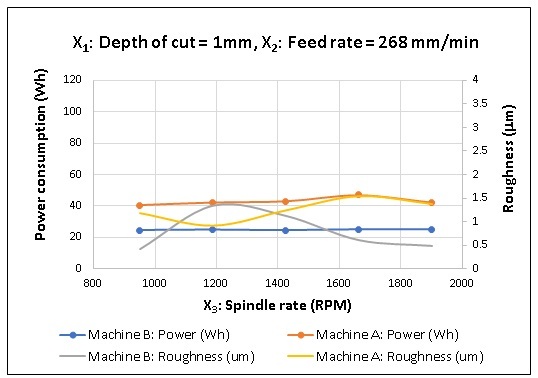}}\\
\subfloat []{\includegraphics[scale=1.4]{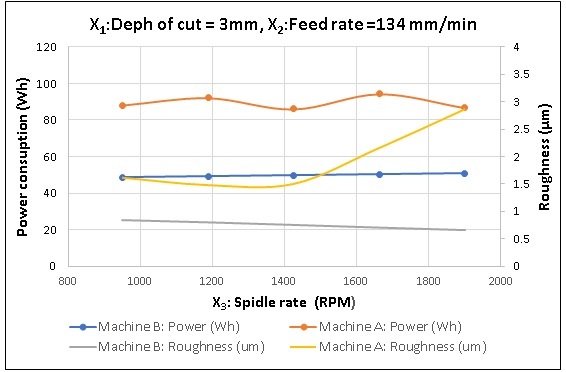}}
\subfloat[]{\includegraphics[scale=1.4]{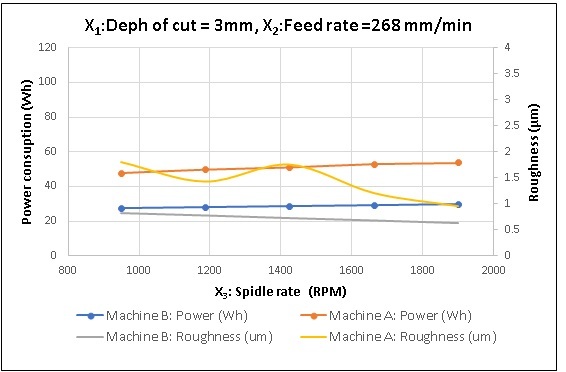}}
\caption{Power consumption and roughness for lower and higher experimental levels at any spindle speed for both machining centers.}
\label{fig_3}
\end{figure}

In this section, the statistical analysis (introduced in section 2.2) of the results related to the experiment described in section 2.1 are presented. However, some descriptive analysis is first shown to give a better understanding of the effect of the three controllable factors ($X_1$ (depth of cut), $X_2$ (feed rate), $X_3$ (spindle rate)) over the two observed responses (power consumption and roughness). 
 Thus, in Figure \ref{fig_3} the patterns of both power consumption and roughness are presented through the different levels of spindle rate while fixing the higher and lower levels for depth of cut ($X_{1}= 1 mm, X_{1}=3  mm$) and feed rate ($X_{2}= 134 mm/min, X_{2}= 268 mm/min$). For both machining centers a significant reduction of the power consumption (up to 45\% of electricity saving) when increasing the feed rate from $134 mm/min$ to $268 mm/min$, for both levels of depth of cut, is clearly observed from the dotted lines in Figure \ref{fig_3}. On the other hand, from the same figure, it can be seen that the roughness has a smooth variability as a function of the spindle rate. In particular, when increasing the spindle rate in machine A, the roughness value increases substantially for the lower level of feed rate. This trend is corroborated in the surface plots presented in the next section. 
 
\begin{figure}[H]
   \centering
\begin{tabular}{cc}
\includegraphics[scale=0.25]{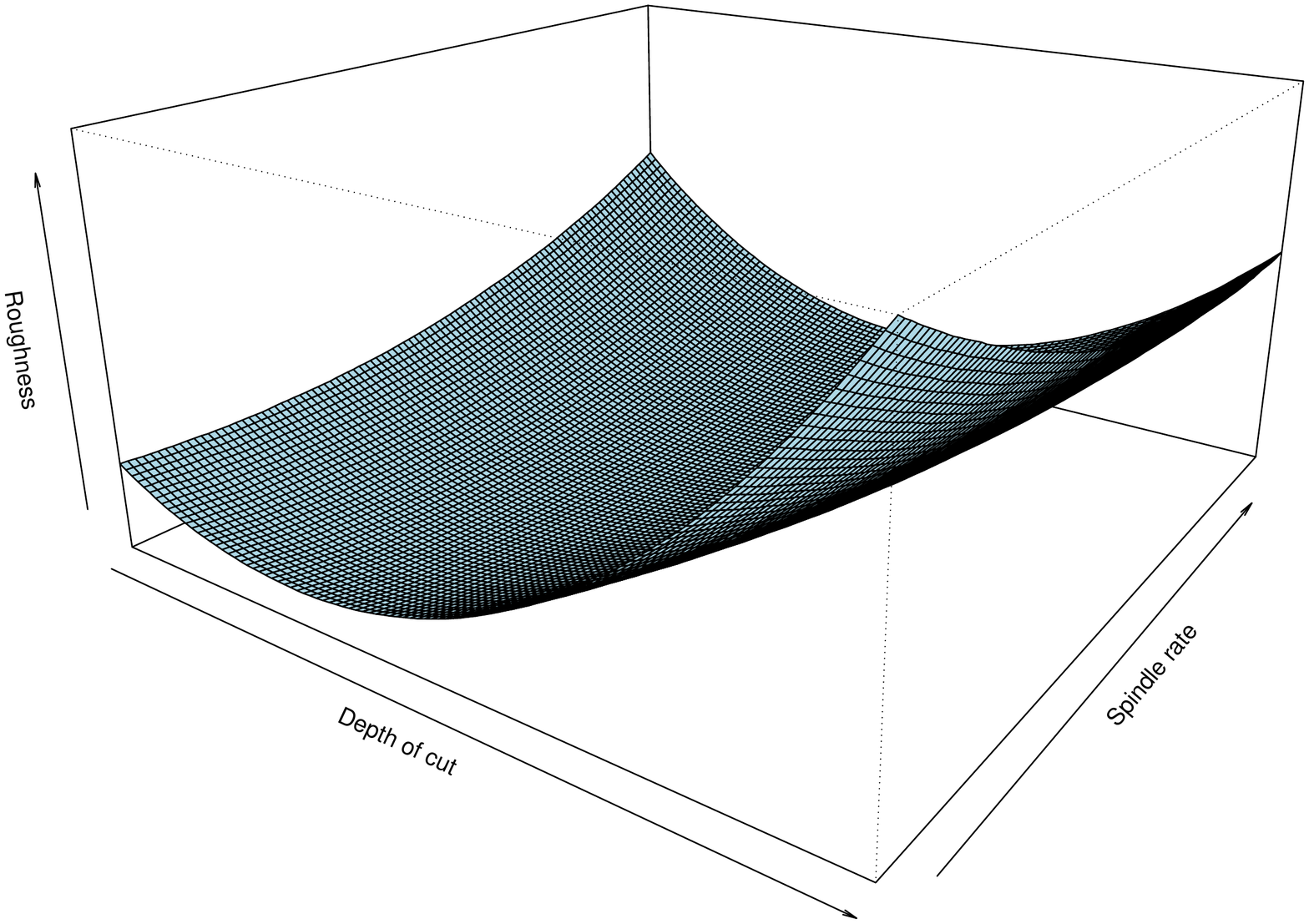}&\includegraphics[scale=0.25]{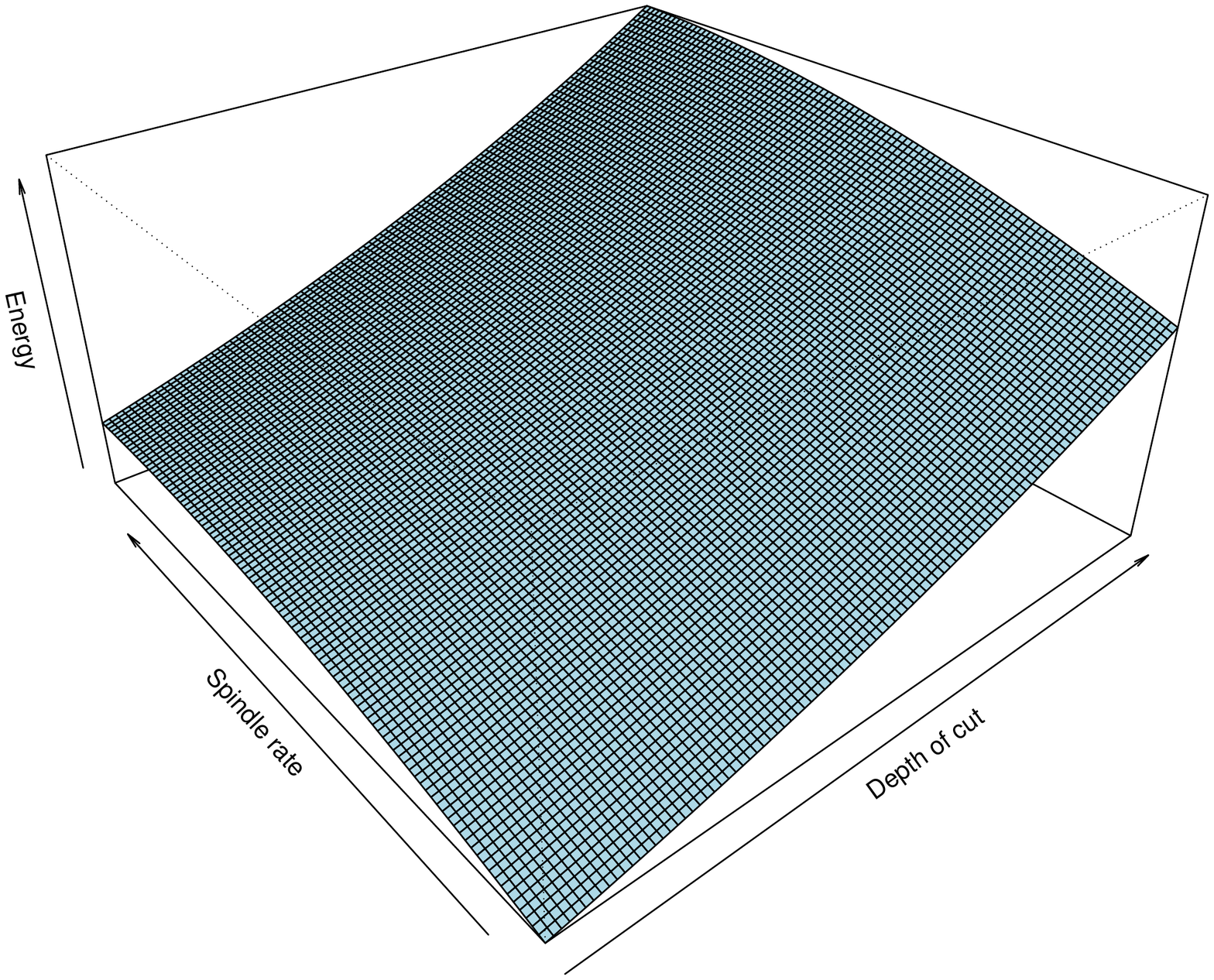}\\(a) & (b)\\
\includegraphics[scale=0.25]{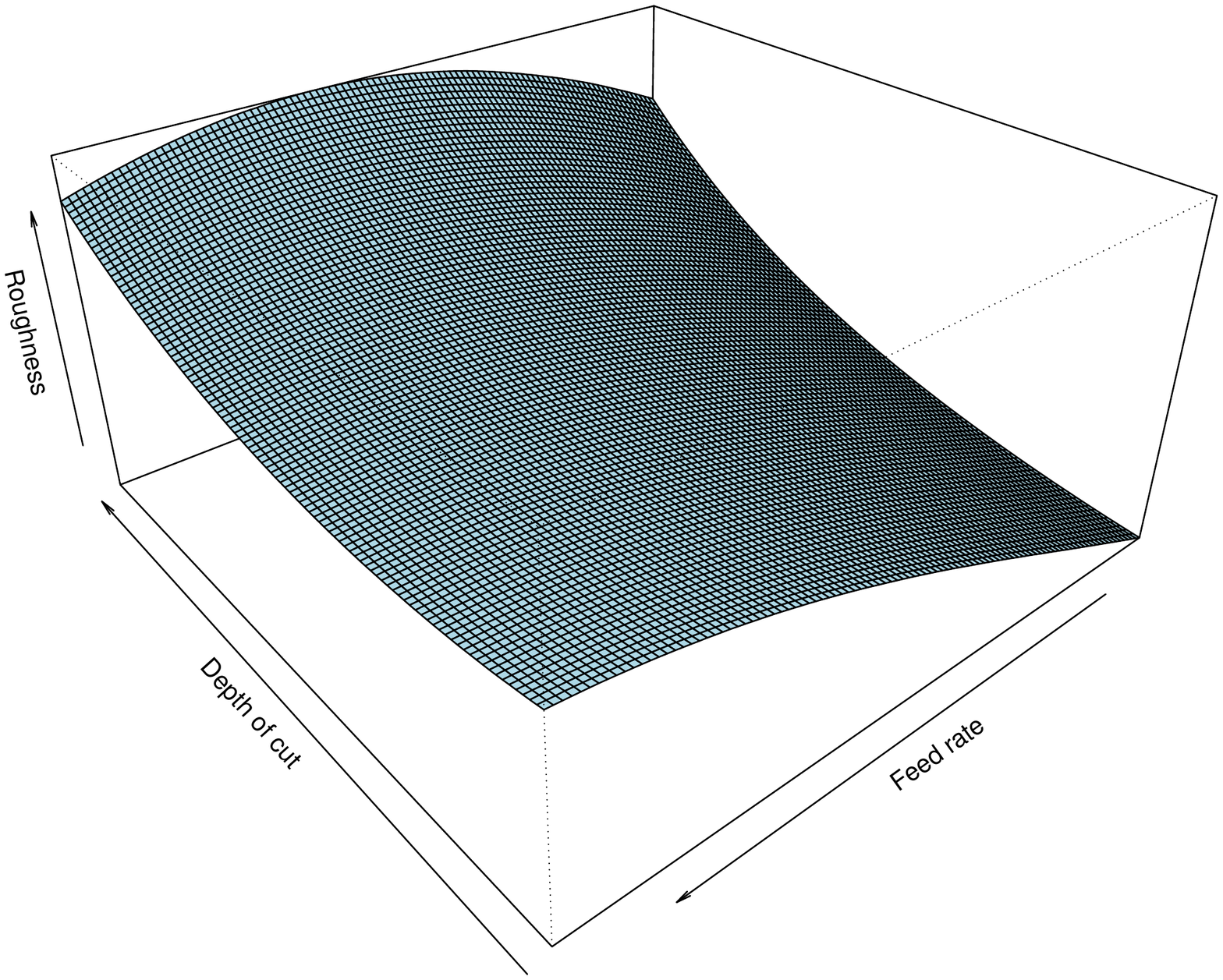}&\includegraphics[scale=0.25]{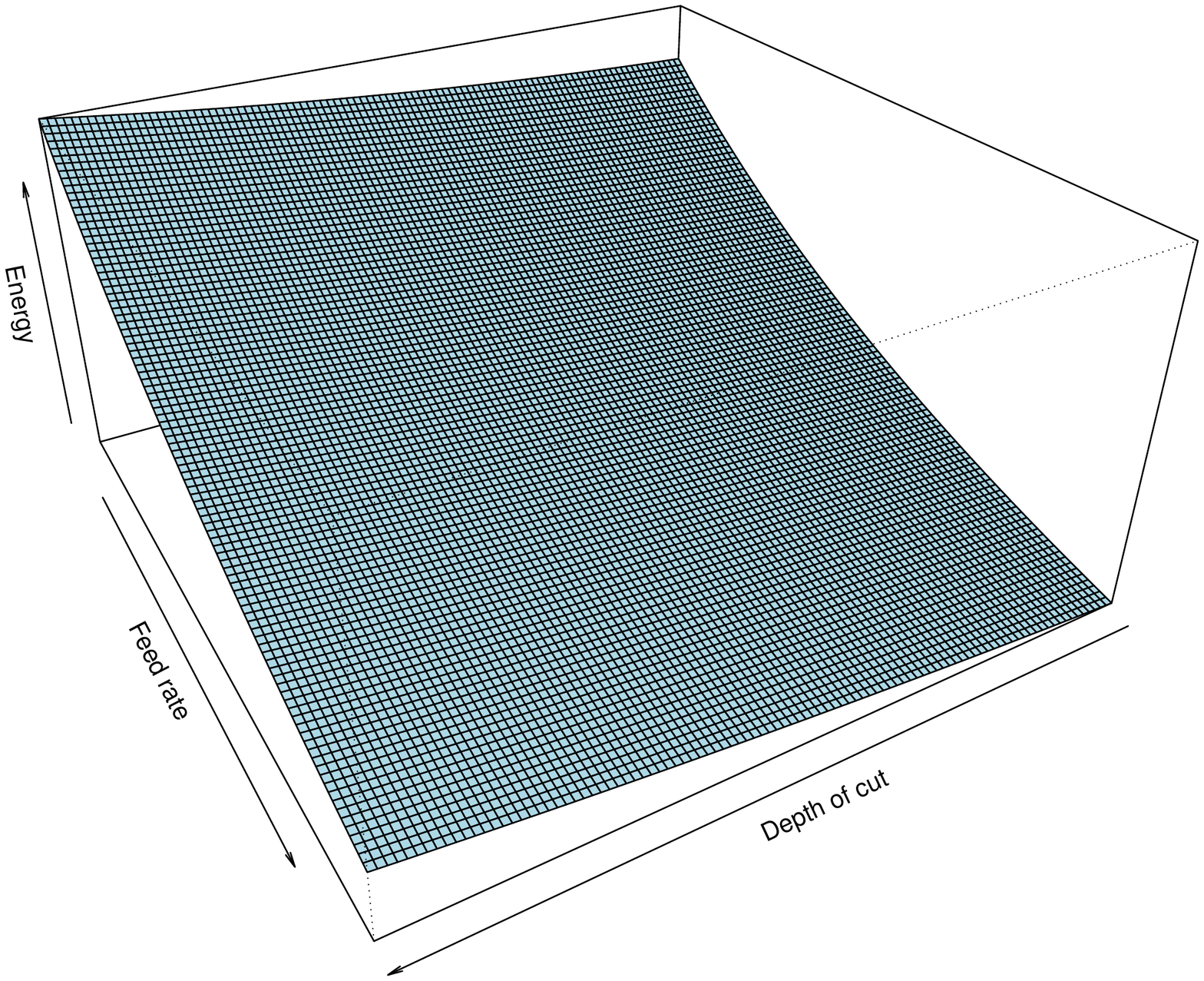}\\(c) & (d)\\
\includegraphics[scale=0.25]{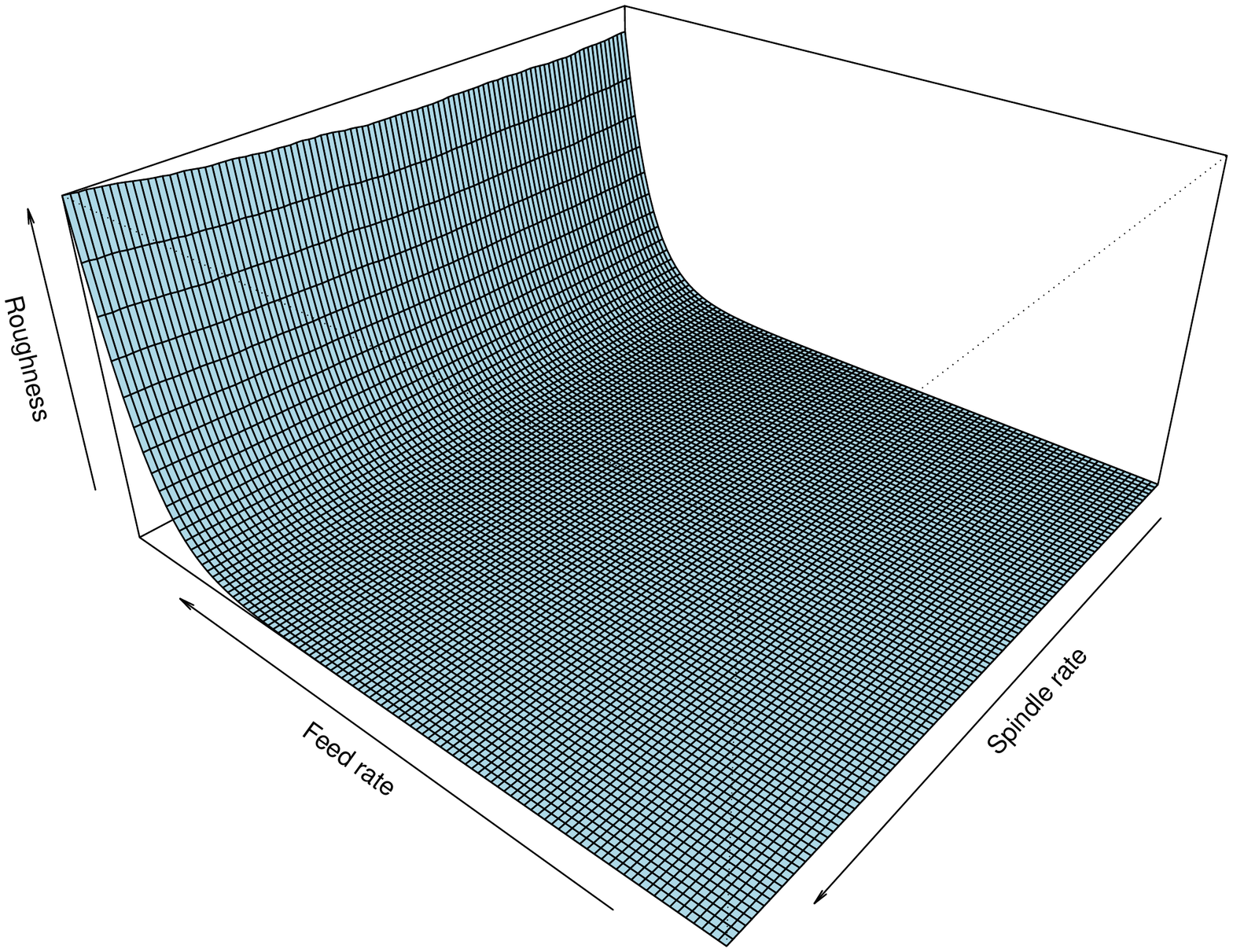}&\includegraphics[scale=0.25]{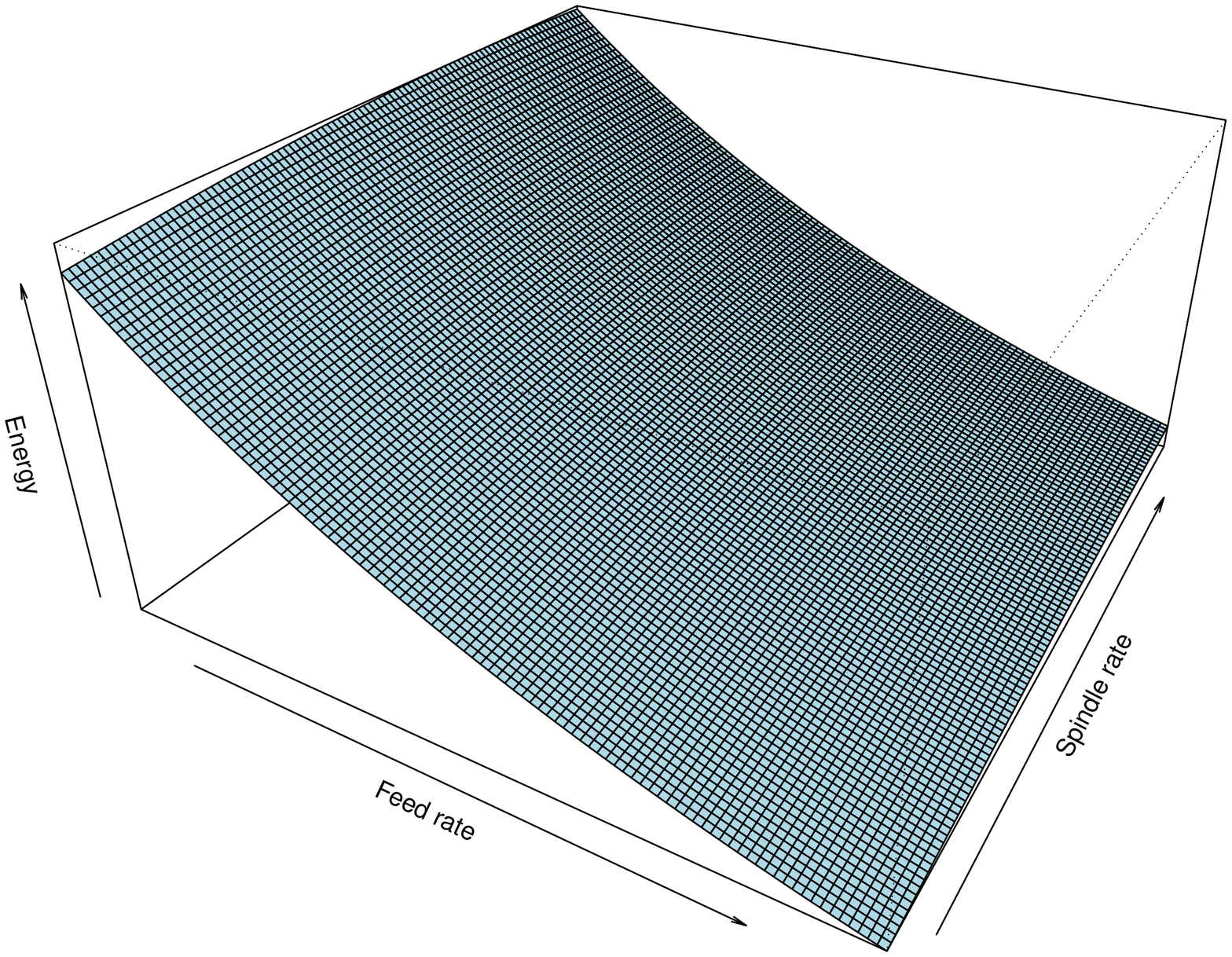}\\(e) & (f)\\
\end{tabular}
\caption{Variation of surface roughness and energy consumption for Leadwell V-40iT machining center (machine A).}
    \label{fig_4} 
\end{figure}

\begin{figure}[H]
   \centering
\begin{tabular}{cc}
\includegraphics[scale=0.25]{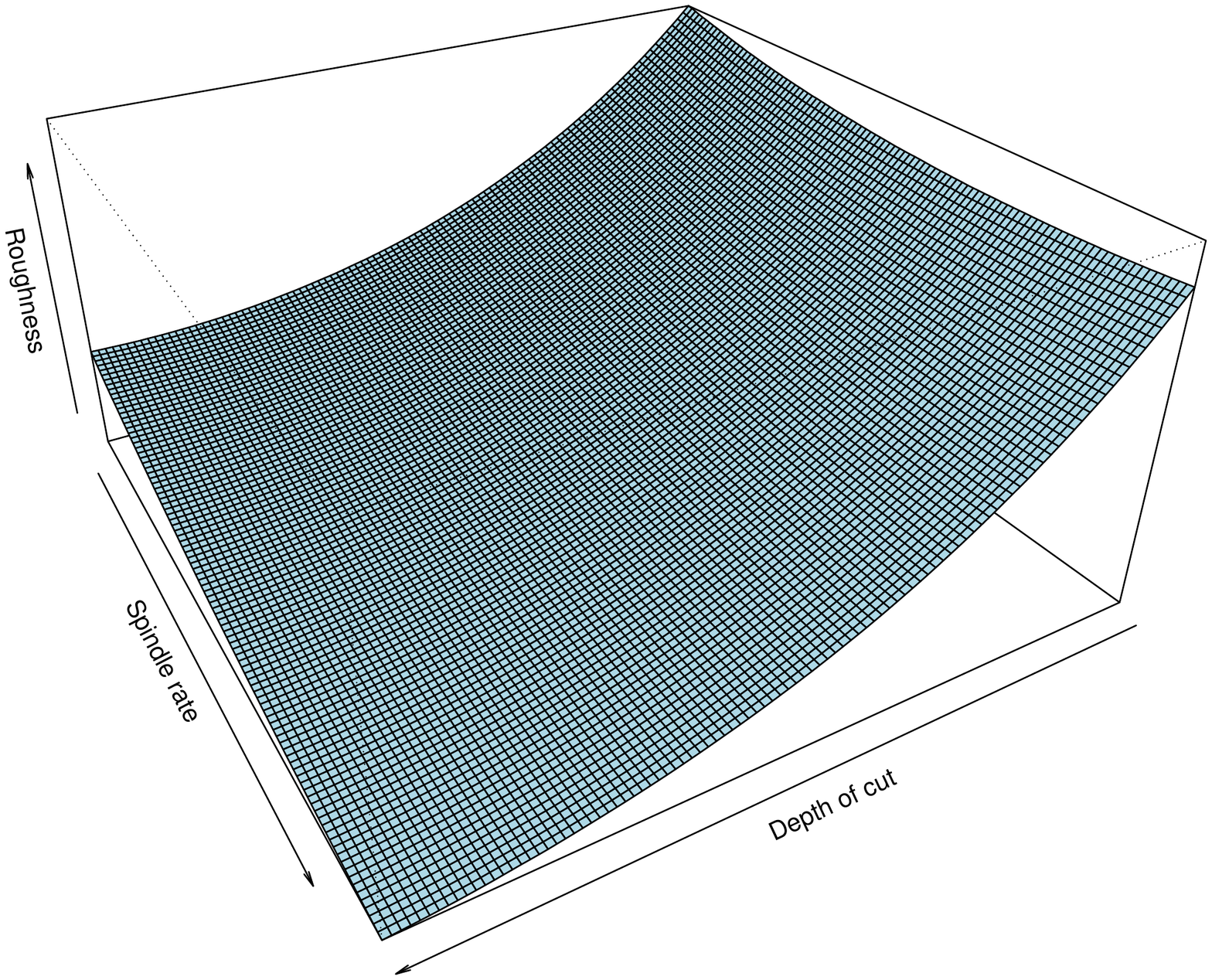}&\includegraphics[scale=0.25]{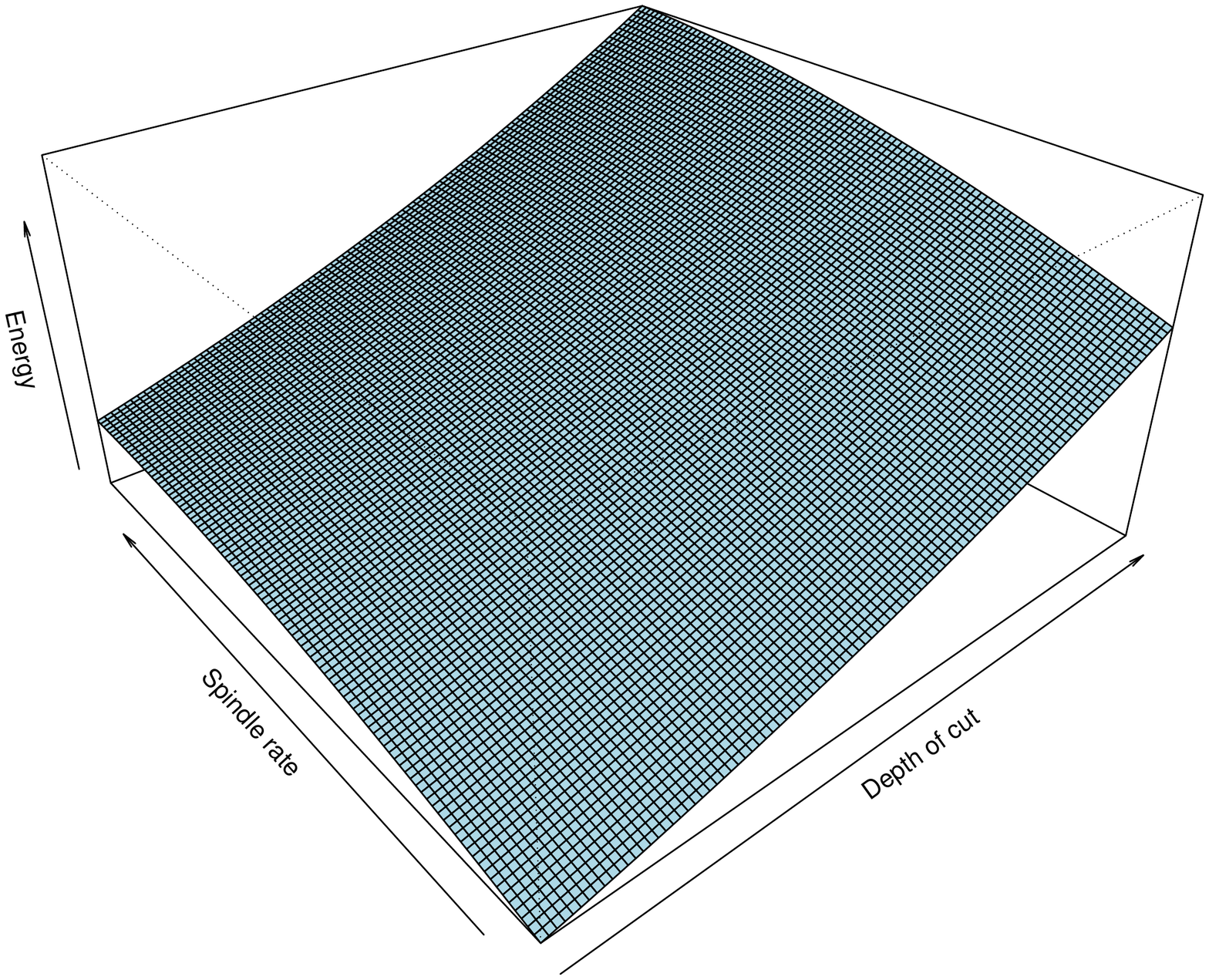}\\(a) & (b)\\
\includegraphics[scale=0.25]{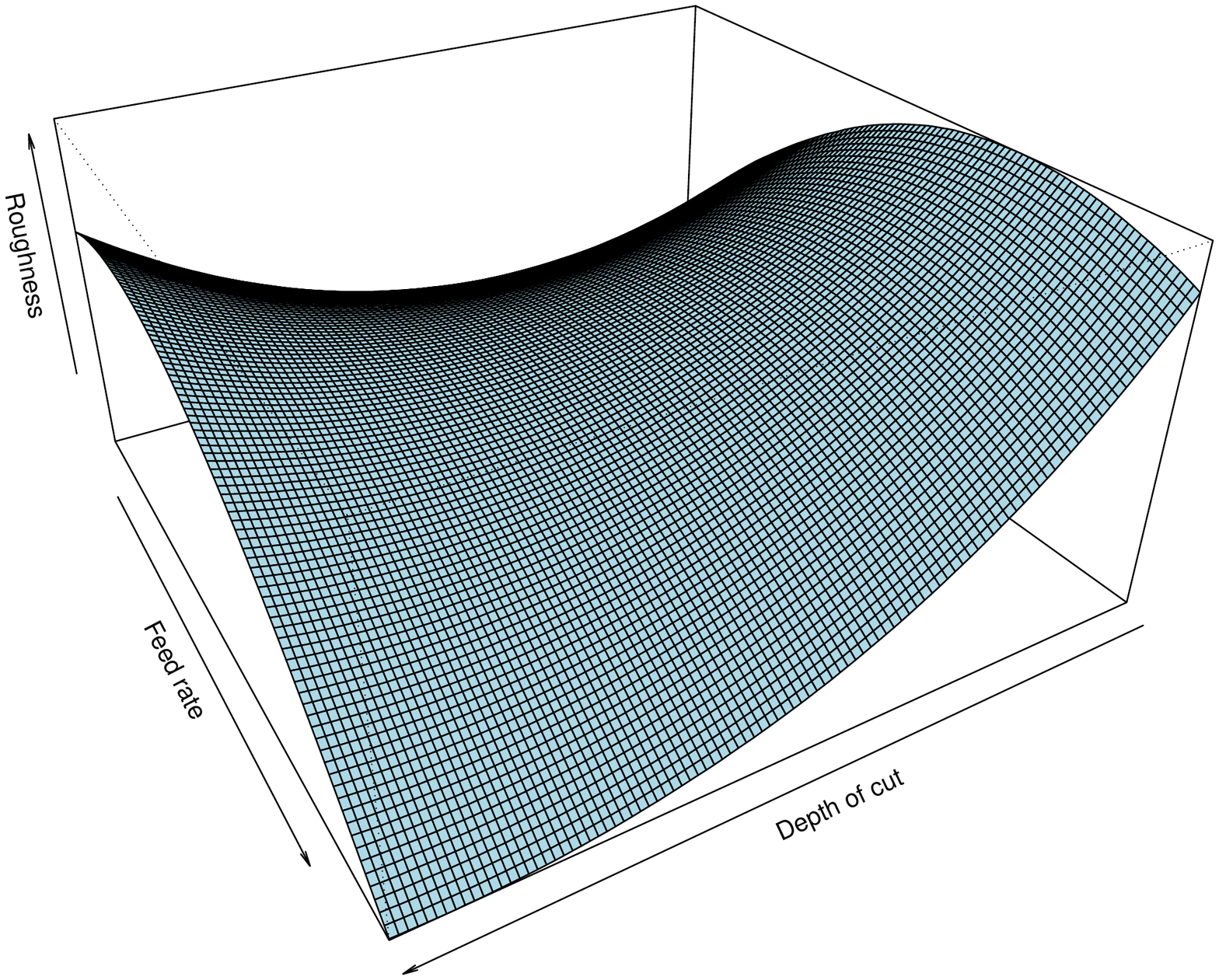}&\includegraphics[scale=0.25]{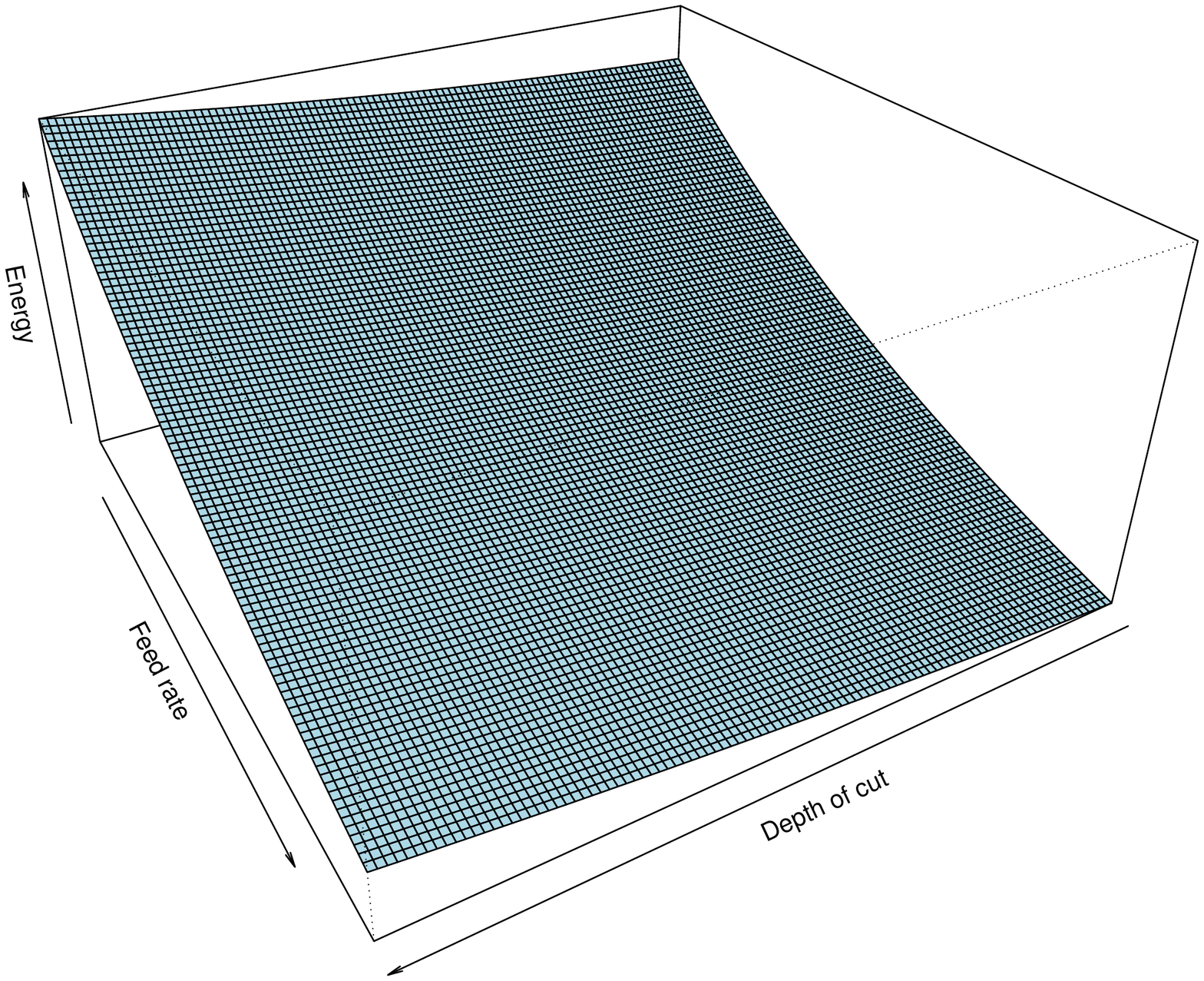}\\(c) & (d)\\
\includegraphics[scale=0.25]{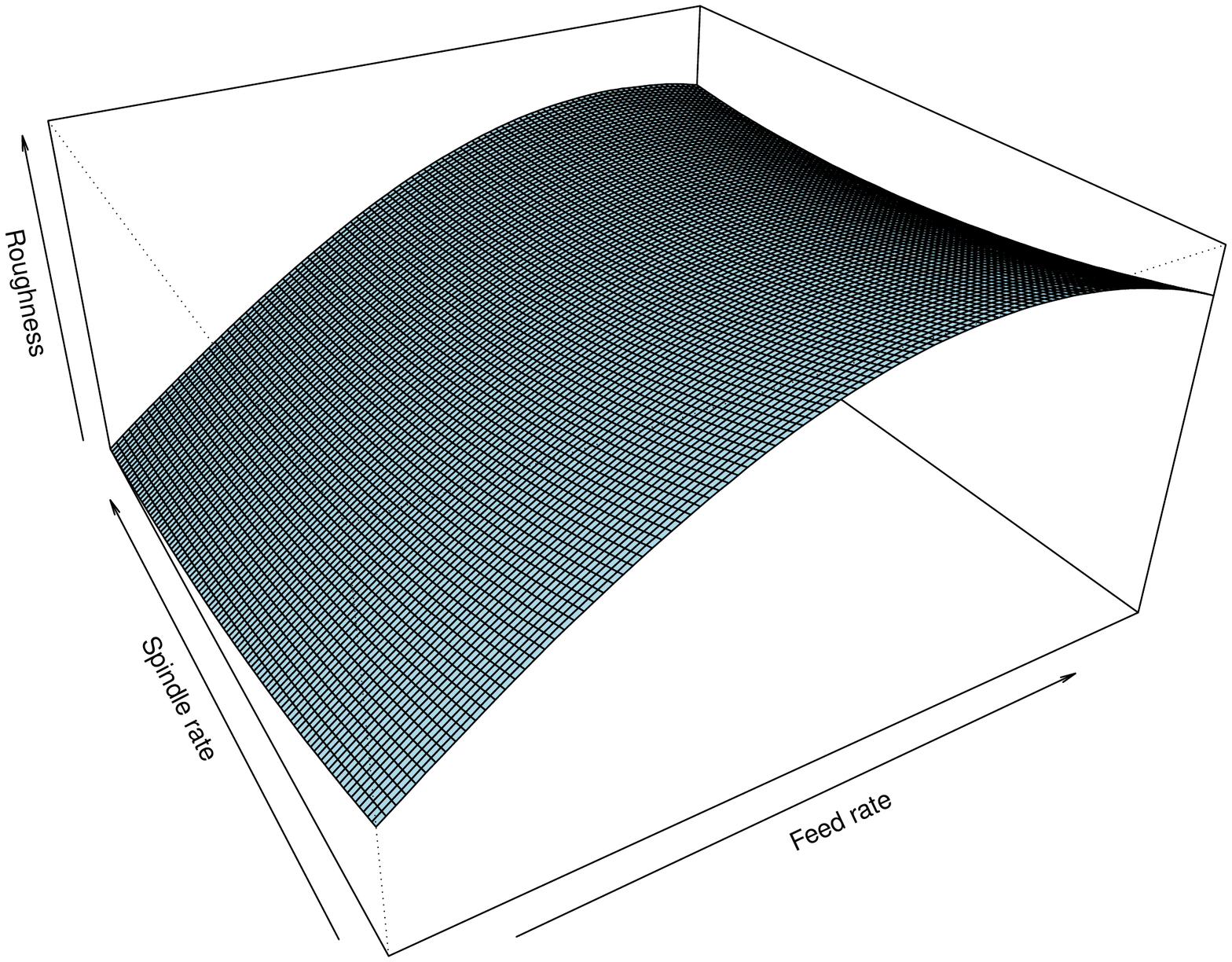}&\includegraphics[scale=0.25]{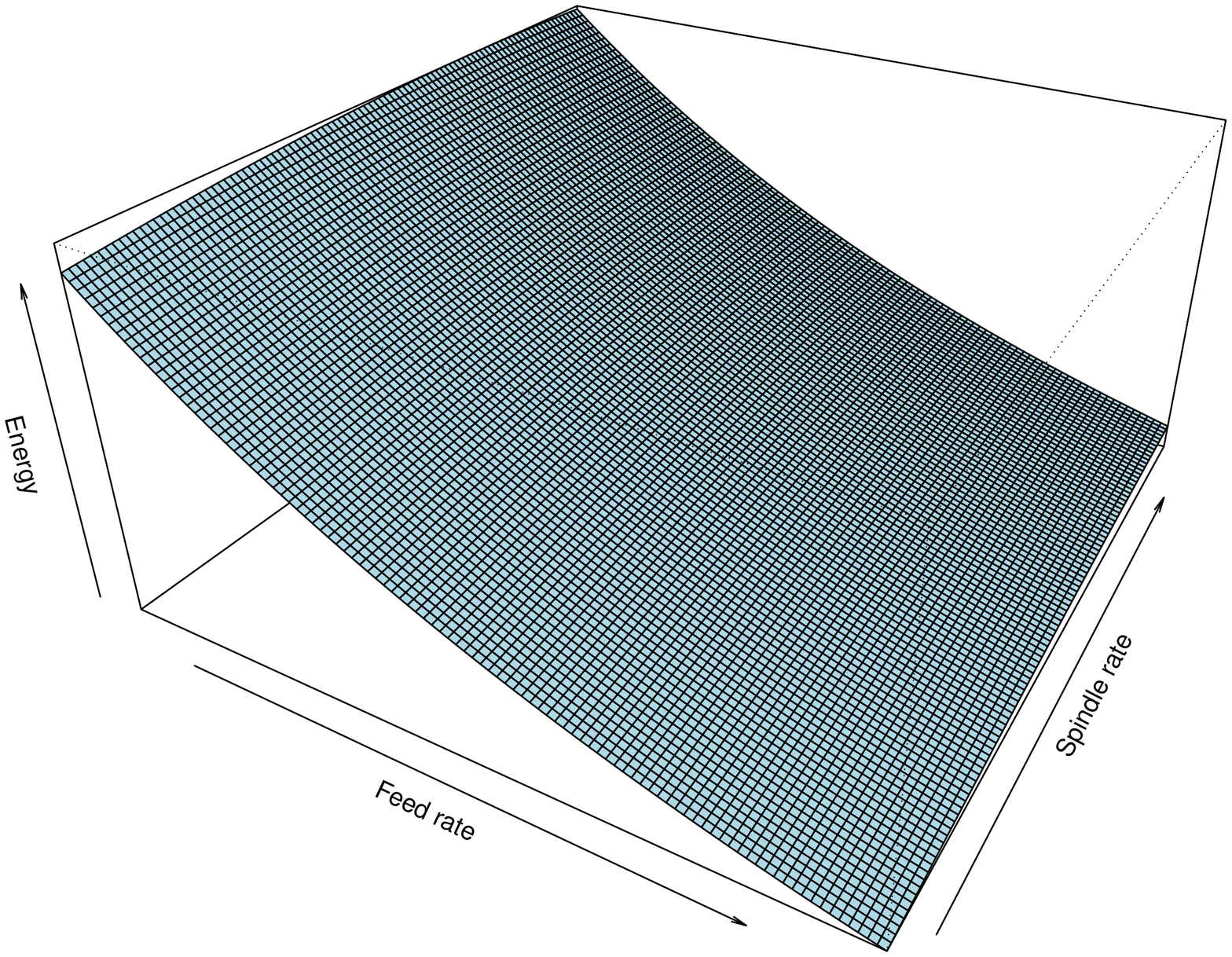}\\(e) & (f)\\
\end{tabular}
\caption{Variation of surface roughness and energy consumption for Haas UMC-750 (machine B)} 
    \label{fig_5} 
\end{figure}

 To perform a more general analysis of the influence of the three factors  $(X_{1}, X_{2}, X_{3})$ on energy consumption and surface finish (roughness), surface plots were obtained from the fitting of model \ref{equ1}. Thus, Figures \ref{fig_4} and \ref{fig_5} show the response surfaces for machining centers A and B, respectively. These graphs are obtained for pairwise factors while fixing the remaining factor on its optimal level according to \ref{funcM}. 
When an increase in the feed rate is set, the outcome for roughness also increases for machines A and B, whereas the spindle rate does not show a significant effect on roughness for both machines A and B. On the other hand, it is interesting to see the influence of the depth of cut on the roughness, which is different for each machining center. In the Leadwell V-40iT machining center (machine A), increasing the depth of cut increases the roughness, reducing the quality of the surface finish (see Figure \ref{fig_4} panels a, c). In contrast, in the Haas UMC-750 (machine B), when the depth of cut increases, the roughness value decreases (see Figure \ref{fig_5} panels a and c). That difference in the effect of the depth of cut on the observed roughness is probably due to differences  in vibration control for each machine. However, the roughness responses are commonly known by their high sensitivity to the operational conditions and the maintenance of the machining center  \cite{Upadhyay2013, Manivel2016, Misaka2020}. 

In terms of power consumption, the relationships    among the three factors are similar in both machining centers (see  Figure \ref{fig_4} panels b, d, f and Figure \ref{fig_5} panels b, d, f). The feed rate which is the factor that determines the duration

for each experiment, shows a direct relationship with the power consumption. That is, when the feed rate increases, the power value increases. On the other hand, the depth of cut shows a slight effect on the power values.

 In Table \ref{tab_3}, the significance analysis for the parameters related to the model (\ref{equ1}) is displayed. Their significance was analysed by using 95\% high-density intervals (HDIs), which are the Bayesian counterparts for 95\% confidence intervals (CIs). Thus, a parameter is classified as significant when its interval excludes the zero value. The parameters related to $X_1$ (depth of cut), $X_2$ (feed rate), $X_3$ (spindle rate), which represent the effects of these factors over the whole observed power consumption and roughness (i.e., without discriminating between machining      centers), have a significant effect on the observed power. For the observed roughness, the results are similar except for $X_3$ (spindle rate), which has a non-significant effect over the whole observed roughness  (as observed in Figures \ref{fig_3}, \ref{fig_4} and \ref{fig_5}). The coefficients related to the interactions $X_1I(m)$, $X_2I(m)$, and $X_2I(m)$ represent the change of the overall effects related to $X_1$, $X_2$, and $X_3$ when going from machining center $A$ to machining center $B$. Those effects are not significant for the observed power, which may mean that the power consumption mechanisms for the two machining centers are not significantly different over the levels observed on $X_1$, $X_2$, and $X_3$. On the other hand, those same effects for the observed roughness are significant (at least for $X_1I(m)$ and $X_2I(m)$), which means that the changes in roughness over the levels observed on $X_1$, $X_2$ are statistically different between the two machining centers $A$ and $B$. For the quadratic terms, only $X_2^2$ is shown to have a significant effect on both responses. The practical interpretation for including nonlinear terms is simply to capture patterns beyond marginal linearity.  In the same vein, the interaction terms $X_1X_2$, $X_1X_3$, and $X_2X_3$ are also included, and among them, only the interaction $X_1X_2$ is shown to have a significant effect on both responses. From this interaction term, one can conclude that the particular effect of $X_1$ (at any level of depth of cut) on both responses will depend on the level set for $X_2$ (feed rate) and vice versa.

 \begin{center}
\begin{table}[H]
\caption{Analysis of significance of the parameters from model (\ref{equ1}) using high density intervals (HDI).}
\begin{center}
\begin{tabular}{ ccccc }
 \hline
          &\multicolumn{4}{c}{\textbf{Response}}\\\cline{2-5}
          &\multicolumn{2}{c}{Roughness}&\multicolumn{2}{c}{Power consumption}\\\cline{2-5}
 \textbf{Variable} & HDI & Significant& HDI & Significant \\ \hline
 $X_1$ & (0.054,0.224) &Yes&(0.031,0.059) & Yes \\ 
 $X_2$ & (0.092,0.260) &Yes&(-0.229-0.201) & Yes \\
 $X_3$ & (-0.112,0.058)&No&(0.002,0.031) & Yes   \\
 $X_1^2$&(-0.012,0.131)&No&(-0.006,0.018) & No   \\
 $X_2^2$&(-0.145,-0.001)&Yes&(0.010,0.034) & Yes  \\
 $X_3^2$&(-0.054,0.088) &No&(-0.020,0.004)  & No   \\
 $I(m)$ &(-0.486,-0.244)&Yes&(-0.571,-0.531) & Yes   \\
 $X_1I(m)$&(-0.335,-0.094)&Yes&(-0.010,0.031)& No       \\
 $X_2I(m)$&(-0.287,-0.046)&Yes&(-0.012,0.028)&  No      \\
 $X_3I(m)$&(-0.169,0.070)&No&(-0.016,0.025)& No\\
 $X_1X_2$&(-0.132,-0.011)&Yes&(0.002,0.022)& Yes     \\
 $X_1X_3$&(-0.085,0.036)&No&(-0.007,0.013)& No    \\
 $X_2X_3$&(-0.043,0.078)&No&(-0.005,0.015)&  No \\
 \hline
 \end{tabular}
\end{center}
\label{tab_3}
\end{table}

\end{center}
 
\begin{figure}[H]
   \centering
\begin{tabular}{cc}
\includegraphics[scale=0.68]{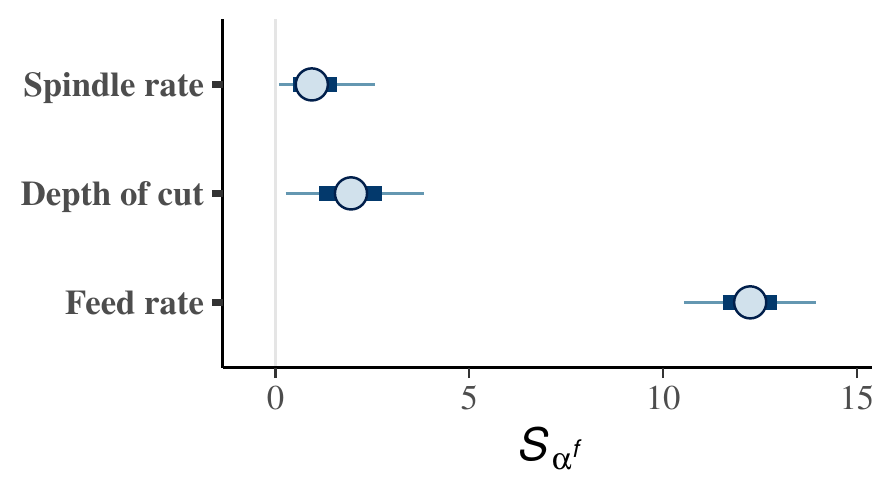}&\includegraphics[scale=0.68]{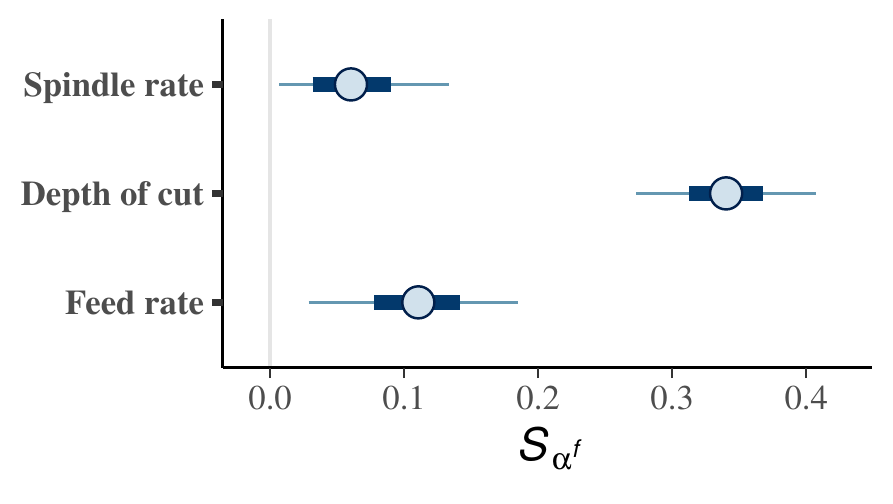}\\(a) Bayesian Anova for power & (b) Bayesian Anova for roughness\\
\end{tabular}
\caption{Bayesian Anova analysis of the model \ref{anova2}. The inner and outer intervals correspond to probability intervals of 50\% and 95\% respectively for $f=\{D:\text{depth of cut}, A:\text{feed rate}, V:\text{spindle rate}\}$.}
\label{fig_6}
\end{figure}

In Figure \ref{fig_6}, the Bayesian ANOVA analysis for the three fixed factors considered in this study are shown. As mentioned in Section 2.4, in contrast to the classic ANOVA analysis, this is a graphical tool that allows identifying the more relevant factors among those considered in the experiment. From this analysis, one can observe that the feed rate is the most relevant factor in explaining the observed variability in power consumed from the two machining centers. In contrast, the most important factor in explaining the observed variability in the roughness is the depth of cut. In addition, for both responses, the spindle rate is the less relevant factor among those analysed in this study in the levels described in Table \ref{tabla_2}.

In Table \ref{tab_4} the results of the optimization process are shown, described in section 2.3. 
The minimum threshold vectors for this process have been defined as $Y^{min}=(40, 0.33)$ and $Y^{min}=(24, 0.28)$  for the machining centers Leadwell V-40iT and Haas UMC-750, respectively. The experimenters define those values, and in this case, they are taken from the minimum values that have been observed in the experiment. The genetic algorithm has taken 20 generations to find the optimal conditions for both machining centers.

 \begin{center}
\begin{table}[H]
\caption{Optimal operational conditions for the two machining centers found by the minimization of the functional $d(E^*)$.}
    \centering
    \begin{tabular}{cccc}\hline
                 &\multicolumn{3}{c}{\textbf{Optimal conditions}}\\\cline{2-4}
      \textbf{Machining center}   & Depth of cut & Feed rate & Spindle rate  \\\hline
       Leadwell V-40iT    & 1 & 268 &950\\
       Haas UMC-750     & 1 & 268 &1111.512\\\hline
    \end{tabular}
        \label{tab_4}
\end{table}
\end{center}

\begin{figure}[H]
\centering
{\includegraphics[scale=0.9]{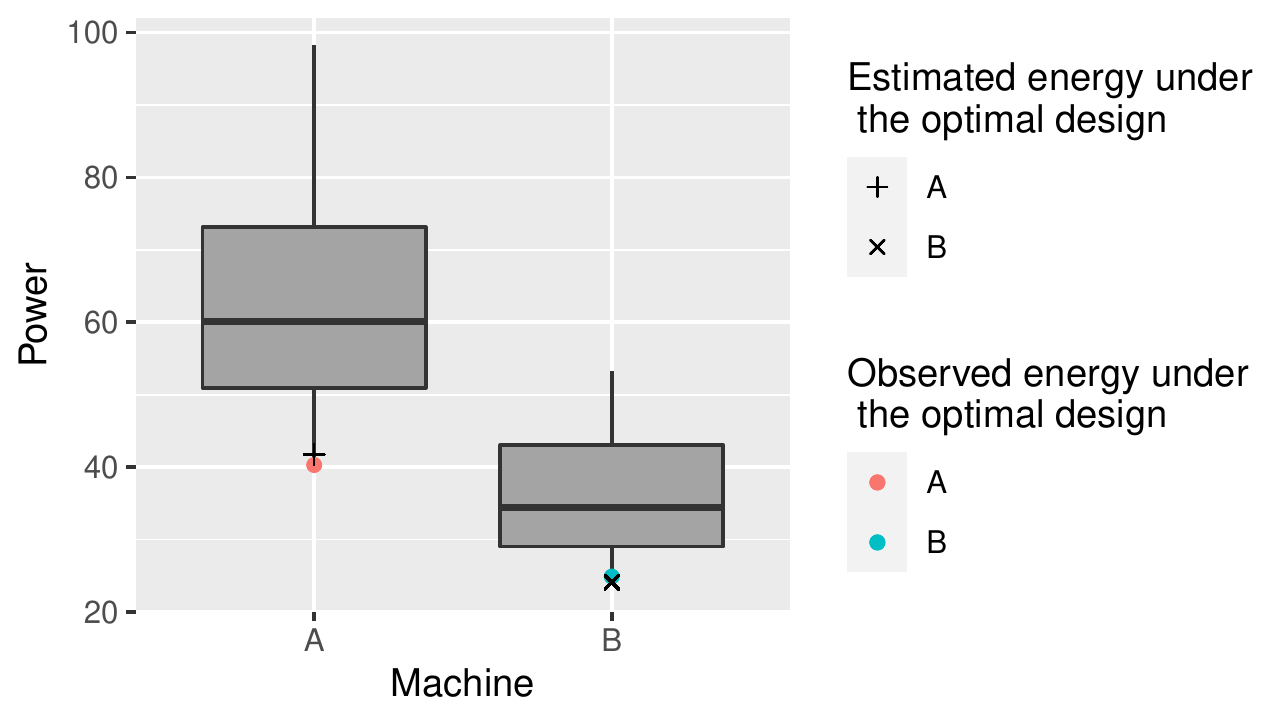}}
\centering
\newline
\subfloat (a) Distribution for the observed power consumption
\newline
\centering
{\includegraphics[scale=0.9]{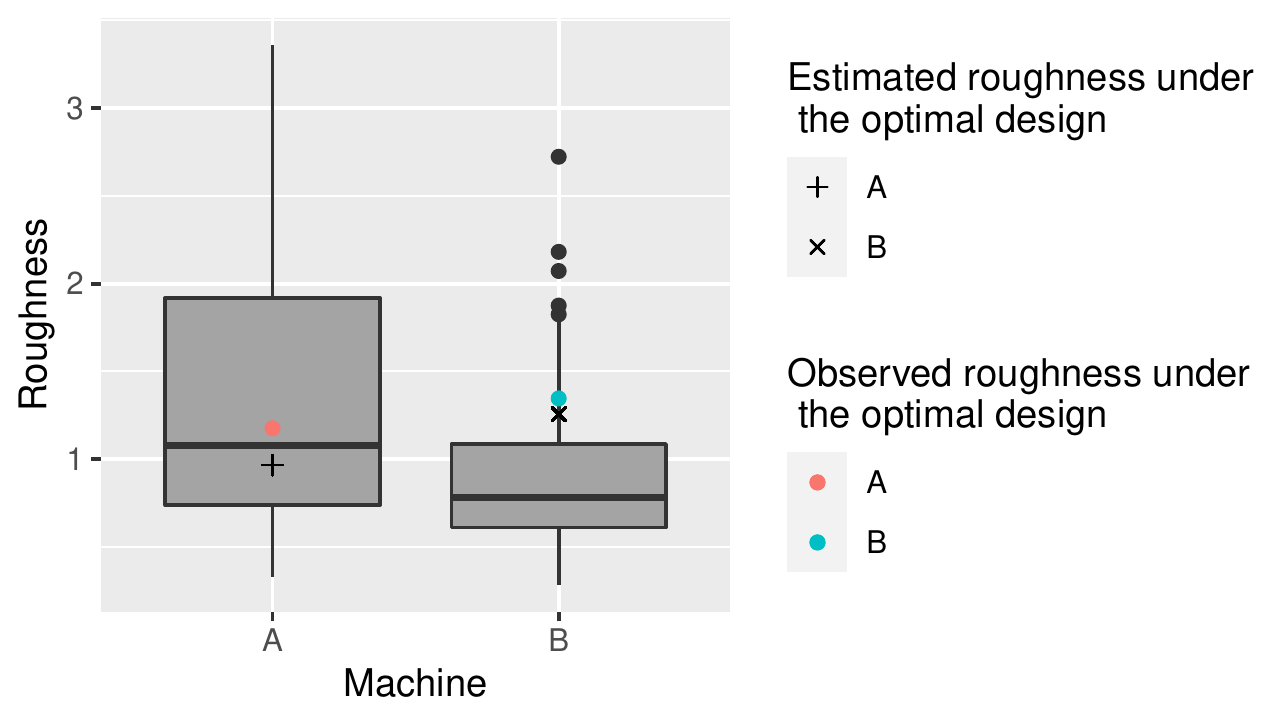}}\\
\centering
\subfloat (b) Distribution for the observed  roughness
\caption{Boxplot for each machining center's observed power and roughness (Machine A: Leadwell V-40iT, Machine B: Haas UMC-750).}
\label{fig_7}
\end{figure}

In Table \ref{tab_5}, Figures \ref{fig_7} and \ref{fig_8} present some important results related to the predictive capability of the model presented in Equation (\ref{equ1}). Thus, in Table \ref{tab_5} the predictions for roughness and power under the optimal operational Conditions $E^*$  are shown. Outstanding performance is presented for the specification defined in model (\ref{equ1}) where low prediction errors are attained. This same analysis has also been performed for the remaining 250 experimental conditions observed in the experiment, and all of them yield prediction errors below 5\%.

\begin{center}
\begin{table}[H]
\caption{Observed and predicted responses under the optimal conditions reported in Table \ref{tab_4}.}
    \centering
    \begin{tabular}{ccc}\hline
     &\multicolumn{2}{c}{\textbf{Machining center}}\\\cline{2-3}
      \textbf{Response}   & Leadwell V-40iT & Haas UMC-750 \\\hline
       \textbf{\textit{Power consumption}} &    & \\
       Observed response     & 40.28 & 24.86 \\
       Predicted response    & 40.66 & 24.07 \\
       Error \%              & 0.01  & 0.03 \\
       \textbf{\textit{Roughness}} &    & \\
     Observed response     & 1.18 & 1.35 \\
       Predicted response    & 1.26 & 0.96 \\
       Error \%              & 0.07  & 0.28 \\ \hline
    \end{tabular}
        \label{tab_5}
\end{table}
\end{center}


In Figure \ref{fig_7} he boxplot analysis is presented for both observed roughness and power consumption along with the location of the predicted and observed responses under the optimal operational conditions presented in Table \ref{tab_4}. From those results, one can conclude that the optimization procedure has identified the optimal design $E^*$ that at least has minimized one of the outcomes (power). In addition, for the other response (roughness), it attained a reasonable level of roughness. One has to bear in mind that the optimization procedure described in Section  2.3 has been built to find the optimal design that simultaneously minimizes both outcomes. Thus, the results presented in Figure \ref{fig_5} ould mean that there are no design or operational conditions which actually minimize both roughness and power consumption at the same time. However, it is possible to find a design that minimizes one outcome (power consumption) while reaching an acceptable result for the other (roughness).In Figure \ref{fig_8} the posterior predictive distributions are presented for both roughness and power consumption under the optimal conditions reported in Table \ref{tab_4}. These results can be interpreted as visual evidence of the predictive performance of model (\ref{equ1}), given that the observed or real responses lie inside of the HDI. The interested reader can find the data and codes to reproduce\begin{figure}[H]
\subfloat (a){\includegraphics[scale=0.65]{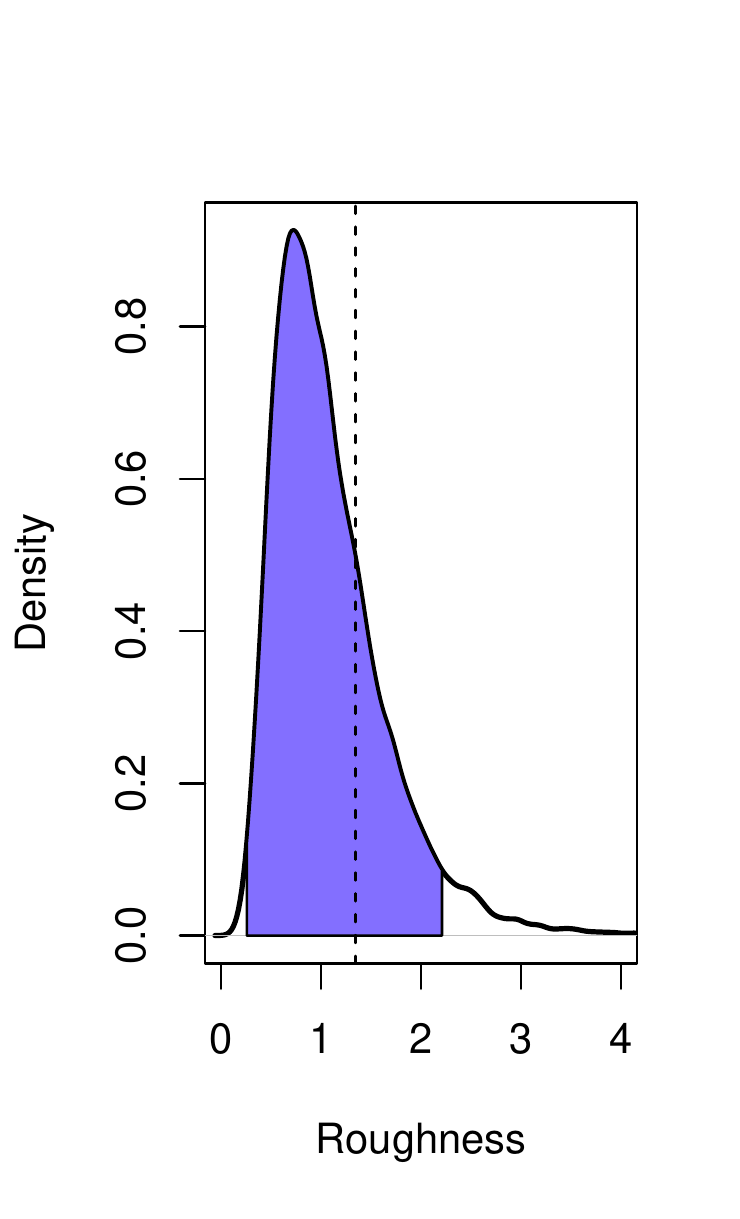}}
\subfloat (b) {\includegraphics[scale=0.65]{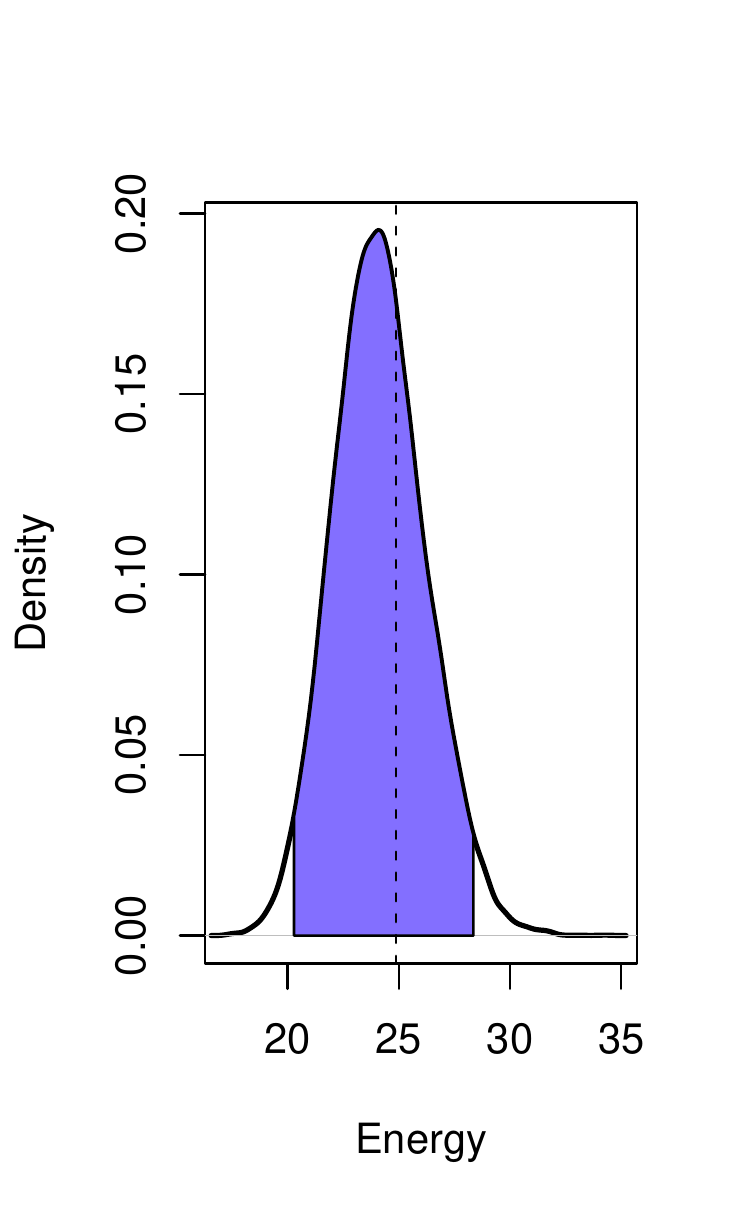}}
\newline
\centering
\subfloat Machining center A\\
\subfloat (c) {\includegraphics[scale=0.65]{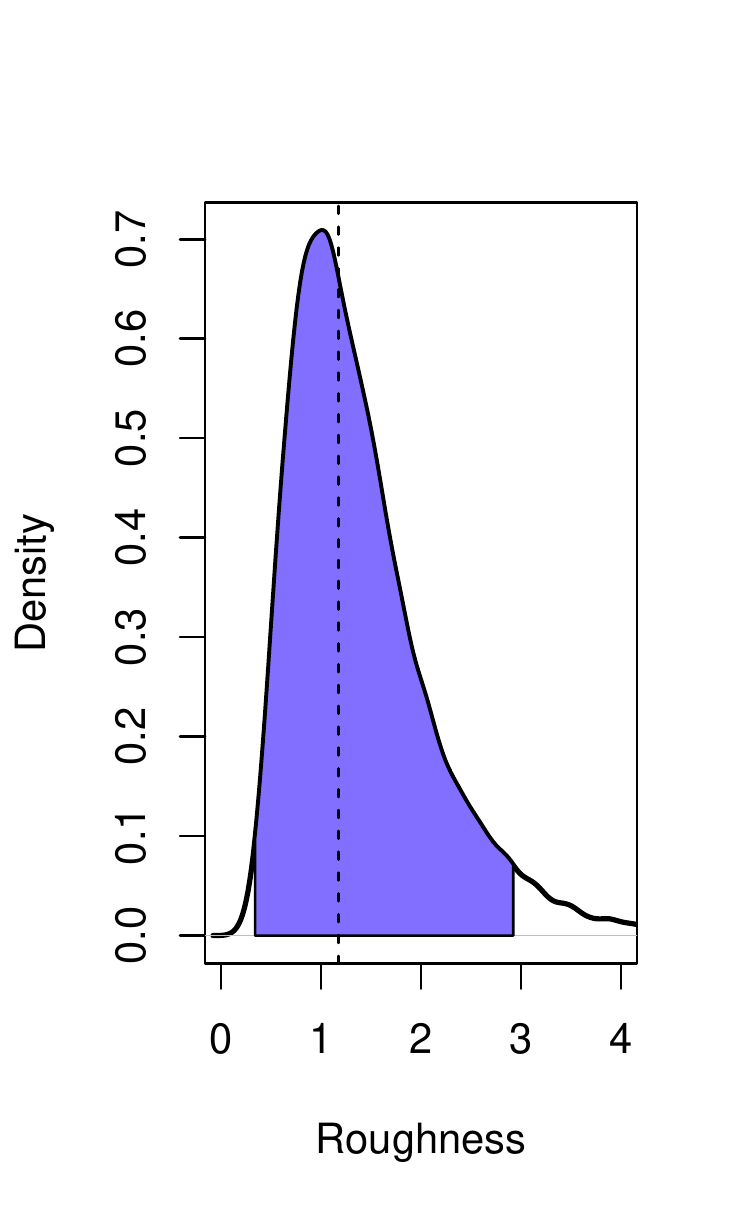}}
\subfloat (d) {\includegraphics[scale=0.65]{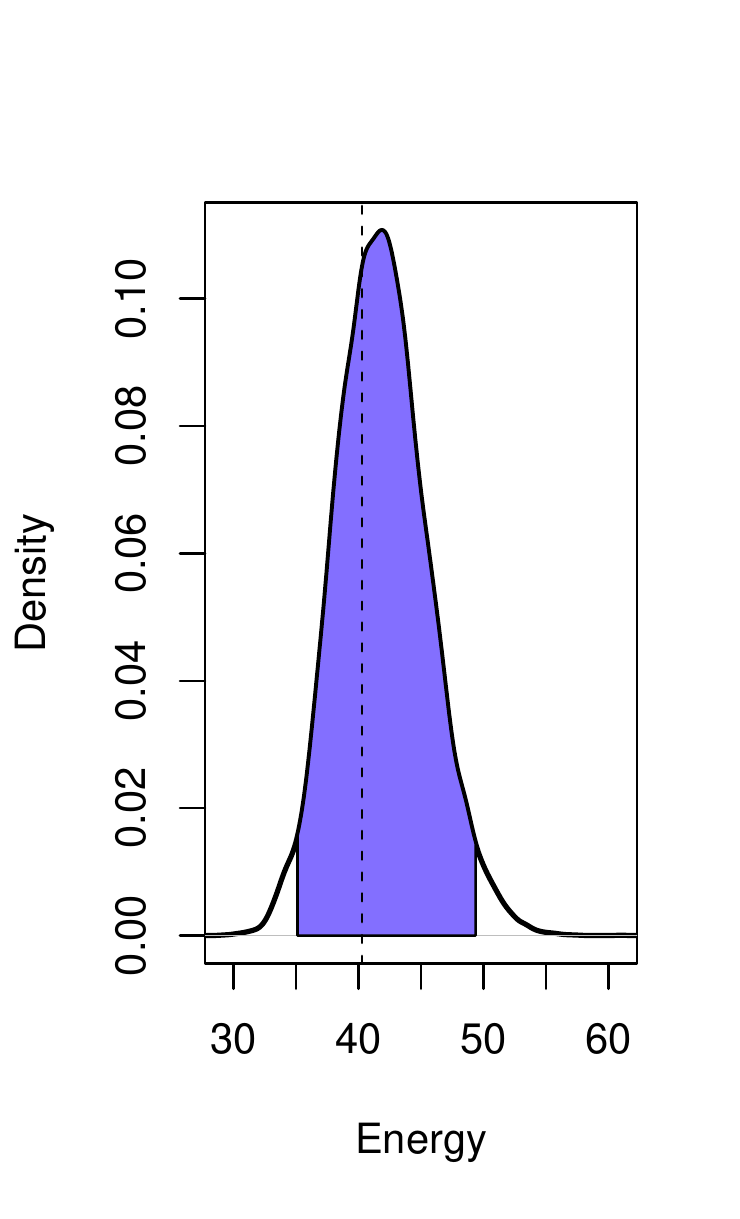}}
\newline
\centering
\subfloat Machining center B
\caption{Densities of the posterior predictive distribution for each machining center evaluated on the optimal design. The dotted lines represent the observed responses reported in table \ref{tab_5} and the colored regions represent the HDI of 95\%.}
\label{fig_8}
\end{figure}the main result from this study at \href{https://github.com/JohnatanLAB/OptimalMachiningCenters}{github.com/JohnatanLAB}.

 \section{Conclusions}
This work presents an experimental study aimed at analysing the influence of operational machining conditions on the performance of two machining      centers. Specifically, the influence of three controllable factors (depth of cut, feed rate, and spindle rate) are studied over the power consumption and surface finish (expressed through the superficial roughness) of two CNC vertical machining centers (The Haas UMC-750 and Leadwell V-40iT).  With a seemingly unrelated regressions (SURs) model, it has been possible to simultaneously model the two observed outcomes of the experiment in one single model, which allows more degrees of freedom, i.e., more data available for the estimation of each parameter. The resulting SUR model has been successfully employed to find the optimal operational conditions by minimizing (\ref{funcM}) using a genetic algorithm. This optimization process suggests that the best performance values for the feed rate ($268 mm/min$) and depth of cut ($1mm$) factors are the same for both machining centers. From the descriptive analysis and Bayesian ANOVA, it is found that the feed rate is a dominant factor in controlling power consumption. Similarly, the  depth of cut reveals a significant relationship with the roughness quality. This work has not considered the performance of the cutting tools as one of the controllable factors. However, to obtain a wide view from an energy efficiency perspective, the geometry and composition of the cutting tools, including the tool path factor, should be considered among the controllable factors in future research.

\section*{acknowledgments}
The authors gratefully acknowledge the financial support provided by  Institución Universitaria Pascual Bravo and Instituto Tecnológico Metropolitano de Medellín.

\section*{Conflicts of Interest}
The authors declare no conflict of interest.

\section*{Author Contributions}

Conceptualization, J.C.J., J.S.R and M.I.A.; methodology, J.S.R, C.A.I.M. E.J.N and J.C.J.; formal analysis, J.C.J., J.S.R., M.I.A. and C.A.I.M. ¸S.; investigation, E.J.N, M.I.A and J.S.R.; resources, E.J.N., M.A.R and J.S.R.; data curation, J.C.J., J.S.R. and M.I.A.; writing—original
draft preparation, J.C.J., J.S.R., and M.I.A.; writing—review and editing, C.A.I.M., J.S.R, J.C.J., and M.A.R; administration,  J.S.R. and M.I.A.; final
revision, All authors. All authors have read and agreed to the published version of the manuscrip

 \bibliographystyle{elsarticle-num} 
\bibliography{elsarticle-template-num}





\end{document}